\documentclass[12pt, amsmath, amssymb]{iopart}

\usepackage{mathrsfs}
\usepackage{graphicx}
\usepackage{subfigure}
\usepackage{amssymb}
\usepackage{epsf,latexsym}
\usepackage{times}
\usepackage{ifpdf}
\ifpdf
\usepackage{epstopdf}   
\fi

\newcommand{\bra}[1]{\langle#1|}
\newcommand{\ket}[1]{|#1\rangle}

\newcommand{\sz}[0]{\ensuremath{\mathbf{\sigma}_z}}
\newcommand{\sx}[0]{\ensuremath{\mathbf{\sigma}_x}}
\newcommand{\sy}[0]{\ensuremath{\mathbf{\sigma}_y}}
\renewcommand{\sp}[0]{\ensuremath{\mathbf{\sigma}_{+}}}
\newcommand{\sm}[0]{\ensuremath{\mathbf{\sigma}_{-}}}

\newcommand{\be}{\begin{equation}}
\newcommand{\ee}{\end{equation}}
\newcommand{\bea}{\begin{eqnarray}}
\newcommand{\eea}{\end{eqnarray}}

\usepackage{color}

\begin{document}

\bibliographystyle{unsrt}

\title{Generating distributed entanglement from electron currents}

\author{Yuting Ping}
\ead{yuting.ping@materials.ox.ac.uk}
\address{Department of Materials, University of Oxford, Parks Road, Oxford, OX1 3PH, UK}

\author{Avinash Kolli}
\address{Department of Physics and Astronomy, University College London, Gower Street, London, WC1E 6BT, UK}

\author{John H. Jefferson}
\address{Department of Physics, Lancaster University, Lancaster, LA1 4YB, UK}

\author{Brendon W. Lovett}
\ead{b.lovett@hw.ac.uk}
\address{School of Engineering and Physical Sciences, Heriot-Watt University, Edinburgh, EH14 4AS, UK}
\address{Department of Materials, University of Oxford, Parks Road, Oxford, OX1 3PH, UK}

\date{\today}


\begin{abstract}
Several recent experiments have demonstrated the viability of a passive device that can generate spin-entangled currents in two separate leads. However, manipulation and measurement of individual flying qubits in a solid state system has yet to be achieved. This is particularly difficult when a macroscopic number of these indistinguishable qubits are present. In order to access such an entangled current resource, we therefore show how to use it to generate distributed, static entanglement. The spatial separation between the entangled static pair can be much higher than that achieved by only exploiting the tunnelling effects between quantum dots. Our device is completely passive, and requires only weak Coulomb interactions between static and flying spins. We show that the entanglement generated is robust to decoherence for large enough currents.
\end{abstract}

\maketitle

\section{Introduction}

Entanglement is an enabling resource for quantum computing (QC). It must be created and consumed in the process of executing any quantum algorithm~\cite{nielsen00}, something which is most obviously apparent in the measurement-based model of quantum computing~\cite{raussendorf01, kok10}. In this picture, entanglement is first generated to build cluster states, before being consumed by single qubit measurements during the execution of an algorithm. The initial entanglement can be generated between distant qubit nodes~\cite{barrett05, bose99, cabrillo99}, each of which can have its own, dedicated, measurement apparatus. Distributed entanglement would also enable secure communication over long distances~\cite{lo99, shor00} and quantum teleportation~\cite{bennett93}.

Devices that generate entangled currents of pairs of electron spins propagating down different leads have been proposed in theoretical work~\cite{saraga03, oliver02, kolli09}, and recent experiments~\cite{hofstetter09, herrmann10} have begun to demonstrate their feasibility~\cite{recher01, recher02, bena02}. However, it is not clear how such an entangled resource could be used for any of the applications discussed above, since the control and measurement of a single flying solid state qubit has yet to be demonstrated experimentally. Furthermore, the macroscopic nature of the currents makes this even more difficult, especially when there are times when spin pairs enter the same lead~\cite{hofstetter09, herrmann10}. In this paper, we will show that it is possible to convert such mobile entanglement to a static form in a completely passive way, in a very simple device -- thus opening up the possibility of quantum information processors based on entangled currents. Notably, our scheme produces a static pair of entangled electron spins that can have a much higher degree of separation (see Sec.~\ref{sec:fd}) than more conventional protocols for entanglement generation for which the separation is limited by quantum tunnelling or similar local interactions~\cite{recher01, recher02, bena02}. 

\begin{figure}[h]
\begin{center}
\includegraphics[width=0.6\columnwidth]{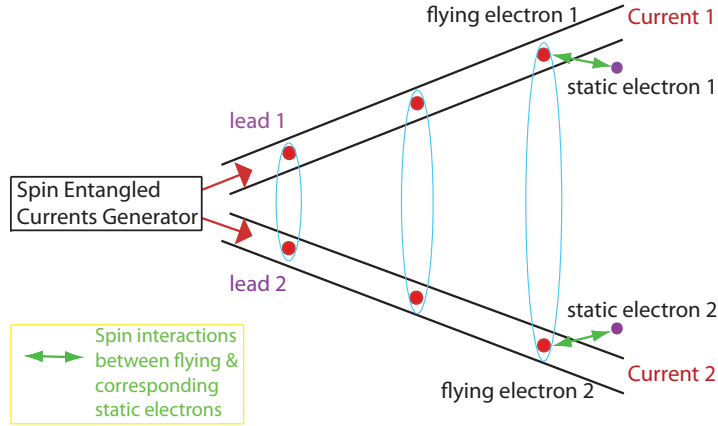}
\caption{Illustration of our entanglement generation device. Two currents, composed of successive electron pairs that are maximally spin-entangled, emerge from a generator and pass down two leads. A static electron is situated downstream of the generator close to each lead, and the static pair is spatially well separated. The mobile spins in each lead interact with corresponding static spins as they pass. Note that the two flying spins of the same Bell pair are not required to, and will normally not, arrive at the sites of interaction with their corresponding static spins at the same time.}
\label{fig:device}
\end{center}
\end{figure}

At the centre of our device is a spin-entangled current generator that outputs entangled pairs of spins $f_i$ down two different leads $i$. Each spin encounters a further, static, spin $s_i$ downstream of the generator and interacts with it, as shown in Fig.~\ref{fig:device}. The generator is based on earlier proposals of a passive device that produces pairs of spin-entangled electrons, with each pair maximally entangled in the singlet Bell state $\ket{S}_{f} =  \frac{1}{\sqrt{2}} (\ket{\uparrow_{f_{1}}\downarrow_{f_{2}}} - \ket{\downarrow_{f_{1}}\uparrow_{f_{2}}})$ ~\cite{saraga03, oliver02, kolli09}. The nature of the static spins is not important, but one possibility is that a single electron is confined in each of the two quantum dots that are fixed close to the leads. Other suitable architectures include endohedral fullerenes in carbon nanotube peapods~\cite{watt08}, carbon nanobud structures~\cite{nasibulin07}, and surface acoustic waves whose minima isolate single electrons~\cite{barnes00}.

\section{Model and Basic Results}

Let us start with an effective Hamiltonian coupling the flying and static spins of the following form: 
\begin{equation}
H_{i} = \frac{g_{i}}{2} (\sx^{s_{i}} \sx^{f_{i}} + \sy^{s_{i}} \sy^{f_{i}}) = g_{i} (\sp^{s_{i}}\sm^{f_{i}} + \sm^{s_{i}} \sp^{f_{i}})
\end{equation}
where the $\sigma_{\pm} = (\sx \pm i\sy)/2$ are the usual Pauli operators. The $g_{i}$ are $XY$ exchange coupling strengths that depend on the separation of the two spins. Each $g_i$ is time dependent since one of the two interacting spins is mobile. The time evolution operator $U_{i} (t)$ for a general state $\ket{\Psi_{i} (t)} = U_{i} (t) \ket{\Psi_{i} (0)}$ of the static-mobile pair $i$, is then
\begin{equation}
U_{i} (t) = \exp\left[- i \theta_i (t) (\sp^{s_{i}}\sm^{f_{i}} + \sm^{s_{i}}\sp^{f_{i}})\right],
\end{equation}
where $\theta_{i} (t) = \int_{0}^{t} g_{i}(t') dt'/\hbar$. $\theta_{i} (t)$ is constant when $[0, t]$ is chosen so that $g_{i} (0)$ and $g_{i} (t)$ are negligible. In the basis $\ket{\uparrow_{s_{i}}\uparrow_{f_{i}}}$, $\ket{\uparrow_{s_{i}}\downarrow_{f_{i}}}$, $\ket{\downarrow_{s_{i}}\uparrow_{f_{i}}}$, $\ket{\downarrow_{s_{i}}\downarrow_{f_{i}}}$,
\begin{equation}
U_{i} =
\left( {\begin{array}{cccc}
1 & 0 & 0 & 0 \\
0 & \cos \theta_{i} & - i \sin \theta_{i} & 0 \\
0 & - i \sin \theta_{i} & \cos \theta_{i} & 0 \\
0 & 0 & 0 & 1 \\
\end{array}}\right).
\label{eq:unitary}
\end{equation}

\subsection{Attractor}

We first consider the case where $\theta_{1} = \theta_{2} = \theta$ and initially the static spins are in the state $\ket{\uparrow_{s_1}\uparrow_{s_2}}$. The starting state of the two static and the first pair of flying spins is then \ $\frac{1}{\sqrt{2}} (\ket{\uparrow_{s_{1}}\uparrow_{f_{1}}} \ket{\uparrow_{s_{2}}\downarrow_{f_{2}}} - \ket{\uparrow_{s_{1}}\downarrow_{f_{1}}}\ket{\uparrow_{s_{2}}\uparrow_{f_{2}}})$. Using the product of 2 unitary operators of the form of Eq.~\ref{eq:unitary}, we find that after interaction the state becomes $\cos \theta \ket{\uparrow\uparrow}_{s} \ket{S}_{f} - i \sin \theta \ket{S}_{s} \ket{\uparrow\uparrow}_{f}$. Since this first flying pair will no longer interact with the static spins nor with the following flying pairs, we can trace out this first flying pair to find the density matrix describing the static pair after the interaction: $\cos^{2} \theta \ket{\uparrow\uparrow}_{s} \bra{\uparrow\uparrow} + \sin^{2} \theta \ket{S}_{s} \bra{S}$.

In order to find the behaviour of our system for multiple passages of flying qubits, we require the following four maps which describe a single interaction event, $U_{1}, U_{2}$:
\begin{eqnarray}
\hspace{-2cm} \ket{\uparrow\uparrow} _{s} \bra{\uparrow\uparrow} &\mapsto& \cos^{2} \theta\ \ket{\uparrow\uparrow}_{s} \bra{\uparrow\uparrow} + \sin^{2} \theta\ \ket{S}_{s} \bra{S}; \nonumber \\
\hspace{-2cm} \ket{\downarrow\downarrow}_{s} \bra{\downarrow\downarrow} &\mapsto& \cos^{2} \theta\ \ket{\downarrow\downarrow}_{s} \bra{\downarrow\downarrow} + \sin^{2} \theta\ \ket{S}_{s} \bra{S}; \nonumber \\
\hspace{-2cm} \ket{S}_{s} \bra{S} &\mapsto& \frac{1}{4} \sin^{2} 2\theta \bigg( \ket{\uparrow\uparrow}_{s} \bra{\uparrow\uparrow} + \ket{\downarrow\downarrow}_{s} \bra{\downarrow\downarrow} \bigg) + \sin^{4} \theta\ \ket{T_{0}}_{s} \bra{T_{0}} +  \cos^{4}\theta\ \ket{S}_{s} \bra{S}; \nonumber \\
\hspace{-2cm} \ket{T_{0}}_{s} \bra{T_{0}} &\mapsto& \ket{T_{0}}_{s} \bra{T_{0}}, 
\label{eq:map1}
\end{eqnarray}
where $\ket{T_{0}} = \frac{1}{\sqrt{2}} (\ket{\uparrow\downarrow} + \ket{\downarrow\uparrow})$, another maximally entangled Bell state. These maps imply that the state $\ket{T_{0}}_{s}$ is an attractor for this process, and by tracking the states of the static spins through multiple passages of the flying spins it is clear that the system will converge towards this maximally entangled attractor; we need not consider any further maps since other static spin states are never accessed. This fixed point is also consistent with spin-invariant scattering of the flying singlets from a static $\ket{T_{0}}$ triplet, as can be verified directly from the Schr\"{o}dinger equation.

\subsection{Convergence} 

\begin{figure}[h]
\begin{center}$
\begin{array}{cc}
  \subfigure[]{\label{fig:theta} \includegraphics[width=2.5in]{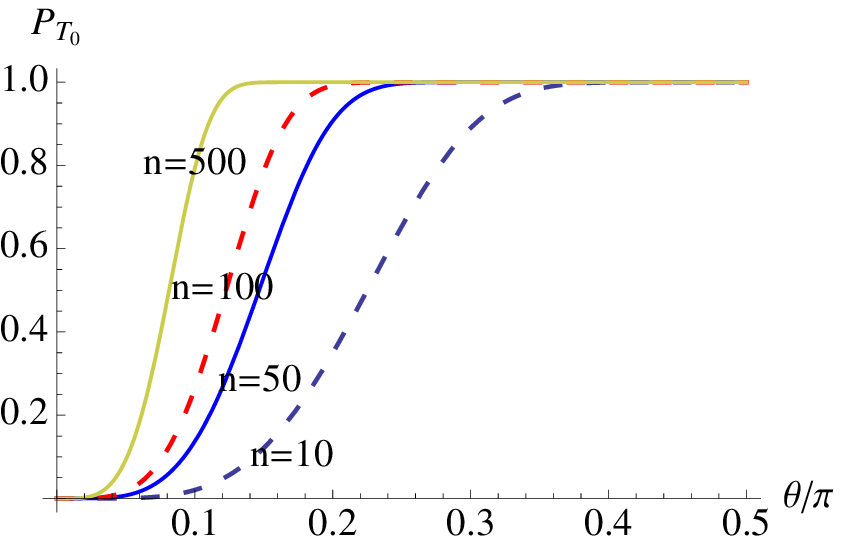}} &
  \subfigure[$ $ $\theta = 0.1$]{\label{fig:rounds0.1} \includegraphics[width=2.5in]{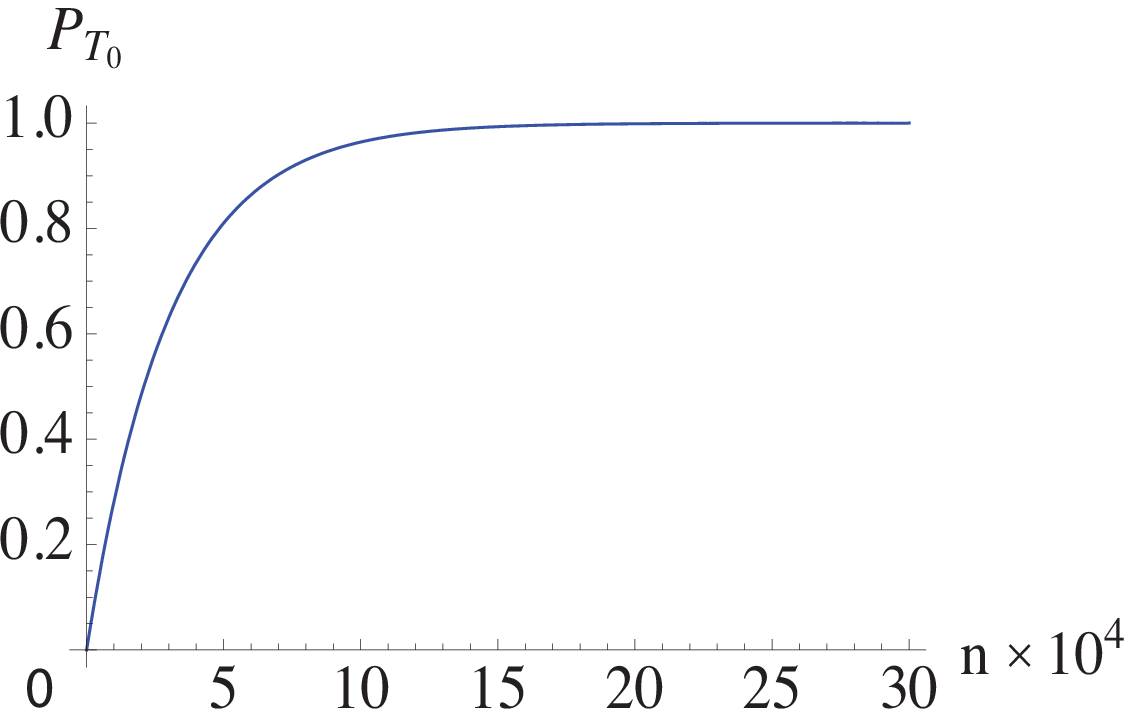}}
\end{array}$
\end{center}
\begin{center}$
\begin{array}{cc}
  \subfigure[$ $ $\theta = 0.03$]{\label{fig:rounds0.03} \includegraphics[width=2.5in]{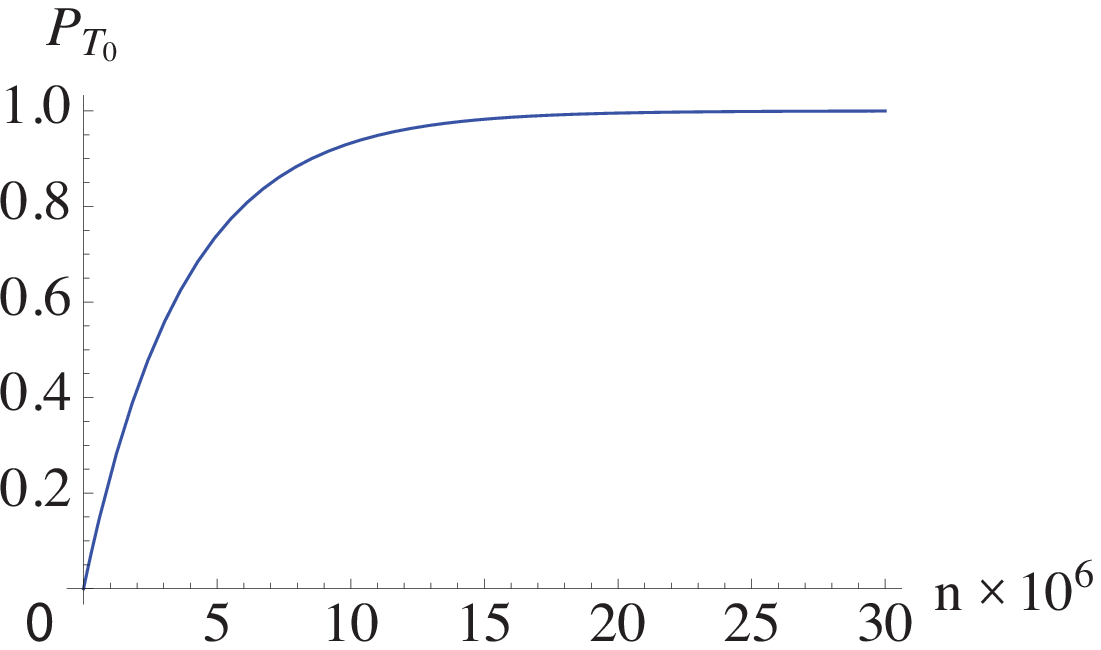}} &
  \subfigure[$ $ $\theta = 0.01$]{\label{fig:rounds0.01} \includegraphics[width=2.5in]{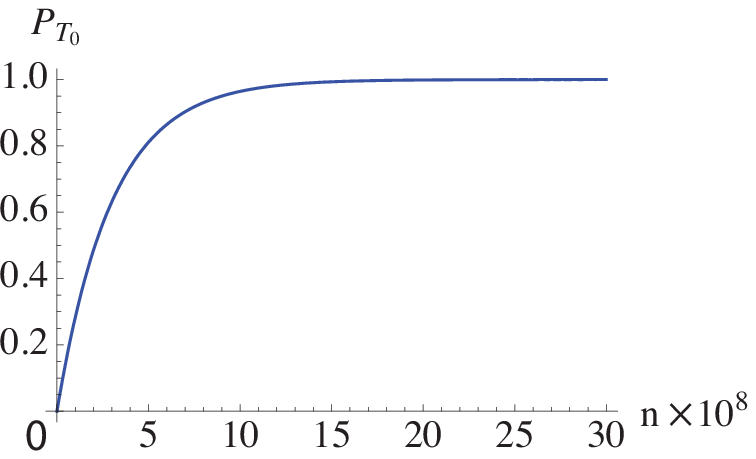}}
\end{array}$
\caption{Plots of the probabilites of obtaining the $\ket{T_{0}}_{s} \bra{T_{0}}$ state (a) against $\frac{\theta}{\pi}$, after $10, 50, 100, 500$ rounds of interactions, respectively; (b) - (d) against the number of rounds of interactions, $n$, for $\theta = 0.1$, $0.03$ and $0.01$, respectively.}
\label{fig:prob}
\end{center}
\end{figure}

We now calculate the probabilities of obtaining the $\ket{T_{0}}_{s}$ state after a number of flying spin passages. The reduced density operator for the static spins after {\it n} iterations is, by definition, $\rho_s^{(n)} = {\bf P}_{n} \centerdot \mathscr{P}$, where $\mathscr{P} \equiv \bigg( \ket{\uparrow\uparrow} _{s} \bra{\uparrow\uparrow}, \ket{\downarrow\downarrow}_{s} \bra{\downarrow\downarrow}, \ket{S}_{s} \bra{S}, \ket{T_{0}}_{s} \bra{T_{0}} \bigg)^T$ is the vector of projection operators for the base states and ${\bf P}_{n} \equiv \big(P_{\uparrow\uparrow}(n),\, P_{\downarrow\downarrow}(n), \,P_S (n),\, P_{T_0} (n)\big)^T$ is the corresponding vector of probabilities. Under the map, Eq.~\ref{eq:map1}, $\mathscr{P} \mapsto L \mathscr{P}$ and hence $\rho_s^{(n)} \mapsto \rho_s^{(n+1)} = {\bf P}_{n} \centerdot (L \mathscr{P}) = (L^T {\bf P}_{n}) \centerdot \mathscr{P}$, i.e., 
\begin{equation}
{\bf P}_n \mapsto {\bf P}_{n+1} = {\bf M_{0}}{\bf P}_n
\label{eq:recur}
\end{equation}
where, by direct substitution,
\begin{equation*}
{\bf M_{0}} \equiv L^T =
\left( {\begin{array}{cccc} 
\cos^{2} \theta & 0 & \frac{1}{4} \sin^{2} 2\theta & 0 \\
0 & \cos^{2} \theta & \frac{1}{4} \sin^{2} 2\theta & 0 \\
\sin^{2} \theta & \sin^{2} \theta & \cos^4\theta & 0 \\
0 & 0 & \sin^4\theta & 1 \\
\end{array}} \right) .
\end{equation*}
Note that the map, Eq.~\ref{eq:recur}, preserves total probability, as it should. Since the matrix ${\bf M_{0}}$ describes the evolution of the state probabilities, its eigenvalues $\lambda_{i}$ must satisfy $|\lambda_{i}| \leqslant 1$. Except when $\theta$ is a multiple of $\pi$, there is always one (and only one) eigenvalue equal to unity, and the corresponding eigenvector is then our attractor state $\ket{T_0}\bra{T_0}$. Using {\it Mathematica} and Eq.~\ref{eq:recur}, we have derived closed analytic expressions for ${\bf M_{0}}^n$ and $P_{T_{0}} (n)$. For the initial conditions ${\bf P}_{0} = (1, 0, 0, 0)^T$, we have
\begin{equation}
1 - P_{T_{0}} (n) = 2^{- 4 n - 1} [(B^n + C^n) - \frac{5 + \cos 2 \theta}{A} (B^n - C^n)]
\end{equation}
where 
\bea
A = \sqrt{2 \cos^{2} \theta (17 + \cos 2\theta)} \nonumber \\
B = 7 + 8 \cos 2\theta + \cos 4\theta - 4 A \sin^{2} \theta \nonumber \\
C = 7 + 8 \cos 2\theta + \cos 4\theta + 4 A \sin^{2} \theta. 
\eea
For large $n$, when $(4 A \sin^{2} \theta)/(7 + 8 \cos 2 \theta + \cos 4 \theta) \in (0, 2)$, i.e., when $\theta \lesssim 0.89$, $C^n$ dominates over $B^n$ and we have 
\begin{equation}
\ln \bigg( 1 - P_{T_{0}} (n) \bigg) \simeq \alpha (\theta)\ n + \beta (\theta)
\label{eq:approx}
\end{equation}
where $\alpha (\theta) = \ln(C/16) \leqslant 0$ and $\beta (\theta) = \ln [(A + 5 + \cos 2 \theta)/(2 A)]$. For $\theta \ll 1$, $(5 + \cos 2 \theta)/A \to 1$, and $\beta (\theta) \to 0$. We plot the probability of obtaining the $\ket{T_{0}}_{s} \bra{T_{0}}$ state against $\theta$, for different values of $n$ in Fig.~\ref{fig:theta}. 

As the number of rounds increases, the range of $\theta$ values for which there is a high probability of obtaining the static $\ket{T_{0}}$ state becomes wider; we plot the probability $P_{T_{0}} (n)$ against the number of rounds, $n$, for some weak coupling strengths in Fig.~\ref{fig:prob}(b)-(d). For the tunnelling rates $\Gamma$ in~\cite{herrmann10}, the time interval $t_{0}$ between successive flying qubits are on the order of $\frac{\hbar}{\Gamma} \simeq 67$ ps, and therefore the time it takes for the static spins to converge to the $\ket{T_{0}}$ state would be 20 $\mu$s, 2 ms and 0.2 s for $\theta = 0.1$, 0.03 and 0.01, respectively. These time intervals $t_{0}$ are at least one order less in~\cite{hofstetter09}, and have the potential of being shortened further. Within the electron spin coherence time ($\geq 200$ $\mu$s) observed in~\cite{bluhm10}, the convergence can occur for $\theta$ as small as 0.03 with the tunnelling rates achieved in~\cite{hofstetter09}. We also point out that molecular systems have the potential for phase shifts $\theta \gtrsim 1$ due to larger exchange interactions arising from nanometer length scales; for example the exchange coupling in a nanotube/fullerene system may be several orders of magnitudes greater than in gated semiconductor devices~\cite{ge08}.

\section{Generalisations} \label{sec:analysis}

Let us now generalize our analysis, to include arbitrary coupling strengths ($\theta_{1} \neq \theta_{2}$), and arbitrary starting states for the static qubits. With the time evolution operators $U_{i}$ defined as in Eq.~\ref{eq:unitary}, we can find a completely positive map $\mathcal{L}_s$ which represents the effect on the static spin density operator $\rho_s$ of a passing flying qubit pair:
\begin{equation}
\rho_{s}^{(k+1)} = \mathcal{L}_{s}[\rho_{s}^{(k)}]  = \Tr_{f}[(U_{1} \bigotimes U_{2}) (\rho_{s}^{(k)} \bigotimes \rho_{f}^{(k)}) (U_{1} \bigotimes U_{2})^{\dagger}]
\label{eq:cpmap}
\end{equation}
where $\Tr_{f}$ denotes the partial trace \cite{nielsen00} over the mobile spins, and  $\rho_{f} = \ket{S}_{f} \bra{S}$.

Eq.~\ref{eq:cpmap} corresponds to a set of 16 recurrence relations for the elements of $\rho_{s}^{(k)}$. When $\theta_{1} = \theta_{2}$, four of these relations decouple from all the others, as we found in our argument earlier. The superoperator $\mathcal{L}_{s}$ is not a linear map, and we thus vectorize the density operator states by listing the entries in the following order as a column vector:  $(\rho_{11} \rho_{12} ... \rho_{14} \rho_{21} ... \rho_{24} \rho_{31} ... ... \rho_{44})^{T} = : \tilde{\rho}$. The map corresponding to the action of the superoperator $\mathcal{L}_{s}$,
\begin{equation}
\tilde{\mathcal{L}}_{s}: \tilde{\rho}_{s}^{(k)} \longmapsto \tilde{\rho}_{s}^{(k+1)}
\end{equation}
is then linear and can be written as a simple $16 \times 16$  matrix {\bf M}, whose entries can be easily calculated using Eq.~\ref{eq:cpmap}. 

We find that {\bf M} always has an eigenvalue $\lambda = 1$, independent of the values of the coupling strengths. The multiplicity of this eigenstate is one, unless $\theta_{1}$ and  $\theta_{2} $ are multiples of $\pi$. The corresponding eigenvector $\tilde{\rho_{1}}$ is then a single attractor state that is independent of the initial configuration of the static spins, which when transformed back to its density matrix form is
\begin{equation}
\rho_{1} = \frac{1}{2(a + b)}
\left( {\begin{array}{cccc}
 a & 0 & 0 & 0  \\
 0 & b & \frac{1}{\sqrt{2}} & 0  \\
 0 & \frac{1}{\sqrt{2}} & b & 0  \\
 0 & 0 & 0 & a  \\
 \end{array} } \right)
\label{eq:evecp}
\end{equation}
where 
\bea
\hspace{- 1.5cm} a = \frac{(\cos \theta_{1} - \cos \theta_{2})^{2} (1 + \cos \theta_{1} \cos \theta_{2}) \csc^{3} \theta_{1} \csc^{3} \theta_{2}}{2 \sqrt{2}} \geq 0, \nonumber \\
\hspace{- 1.5cm}b = \frac{\csc \theta_{1} \csc \theta_{2} (\csc^{2} \theta_{1} + \csc^{2} \theta_{2}) - \cot \theta_{1} \cot \theta_{2} (2 + \csc^{2} \theta_{1} + \csc^{2} \theta_{2})}{2 \sqrt{2}} \geq \frac{1}{\sqrt{2}}.
\eea  
When $\theta_{1} = \theta_{2}$, we have $a = 0$ and $b = \frac{1}{\sqrt{2}}$, and $\rho_{1}$ reduces the $\ket{T_{0}} \bra{T_{0}}$ state as in the simple case. When $\theta_{1}, \theta_{2} \in (0, \pi /2]$, all the other eigenvalues are numerically in the range $(-1,1)$, and their corresponding eigenvectors, when back in matrix form, all have trace zero and hence do not correspond to physical density operators. This has to be the case, as can be seen from the following argument. We can express any density state as $\rho = \sum_{i=1}^{16} a_{i} \rho_{i}$, since the set of eigenvectors $\tilde{\rho_{i}}$ (vectorized $\rho_{i}$) form a basis for {\bf M}. The coefficients $a_{i}$ can take any values, so long as $\Tr(\rho) = 1$ and $\rho$ is positive semi-definite and Hermitian. Hence, $\mathcal{L}_{s} [\rho] = \sum_{i=1}^{16} a_{i} \lambda_{i} \rho_{i}$, or more generally $\mathcal{L}_{s}^{n} [\rho] = \sum_{i=1}^{16} a_{i} \lambda_{i}^{n} \rho_{i}$, which also has trace 1 as a density matrix. As $n \rightarrow \infty$, $\lambda_{i}^{n} \rightarrow 0$ $\forall |\lambda_{i}| < 1$ which is when $i = 2, 3, ..., 16$, and thus $\mathcal{L}_{s}^{n} [\rho] \rightarrow a_{1} \rho_{1}$. So, $\Tr(\mathcal{L}_{s}^{n} [\rho]) \rightarrow a_{1}$ since we defined $\Tr(\rho_{1}) = 1$, and this requires $a_{1} = 1$. We thus obtain $\rho = \rho_{1} + \Sigma_{i=2}^{16} a_{i} \rho_{i}$, and the trace requirement result in $\Sigma_{i=2}^{16} a_{i} \Tr(\rho_{i}) = 0$, which holds for various combinations of $a_{i}$'s. This can only be true when $\Tr(\rho_{i}) = 0$ $\forall i = 2, 3, ..., 16$. $\square$. 

\begin{figure}[h]
\begin{center}$
\begin{array}{cc}
  \subfigure[Fidelity]{\label{fig:fidenta} \includegraphics[width=2.2in]{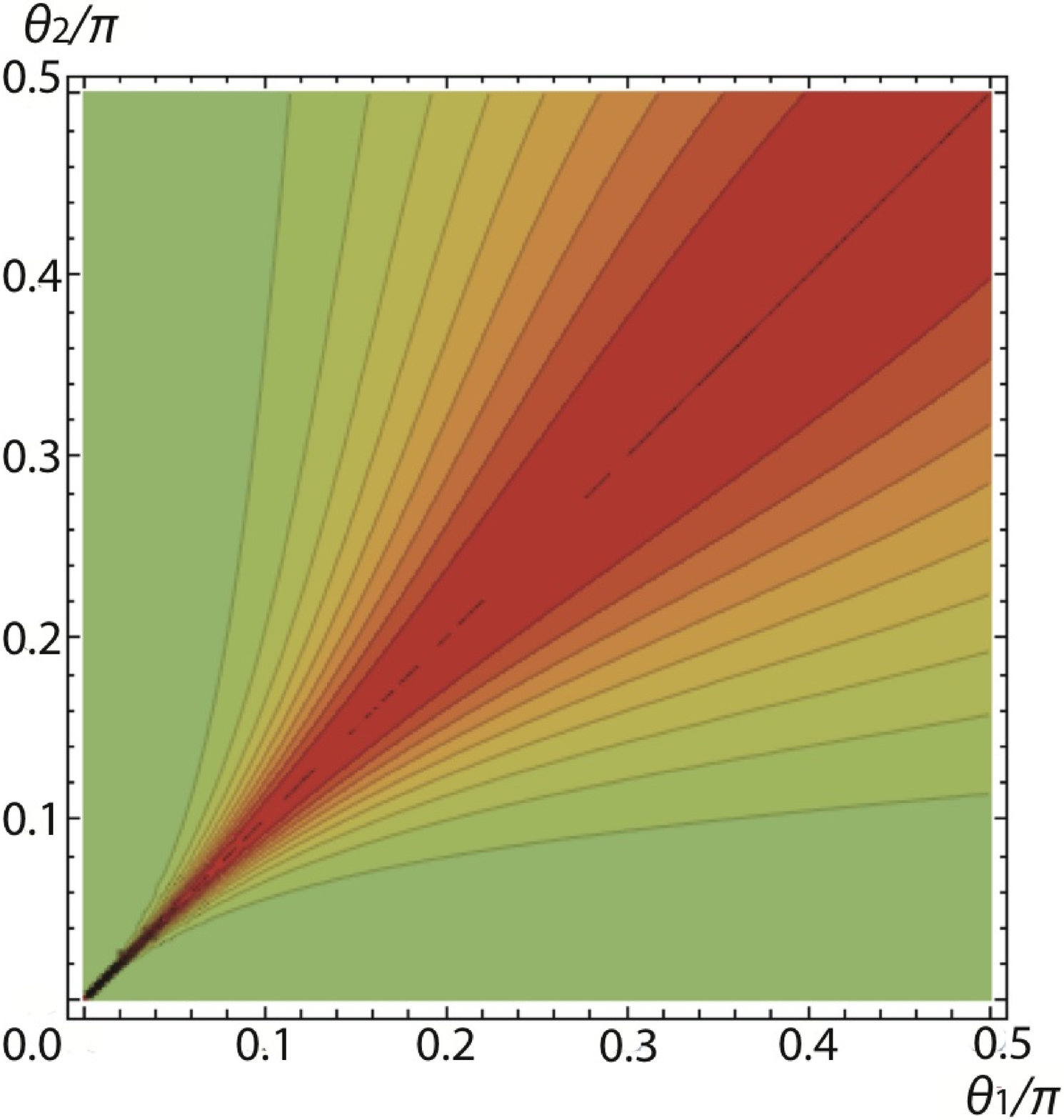}} &
  \subfigure[$E_{F}$]{\label{fig:fidentb} \includegraphics[width=2.2in]{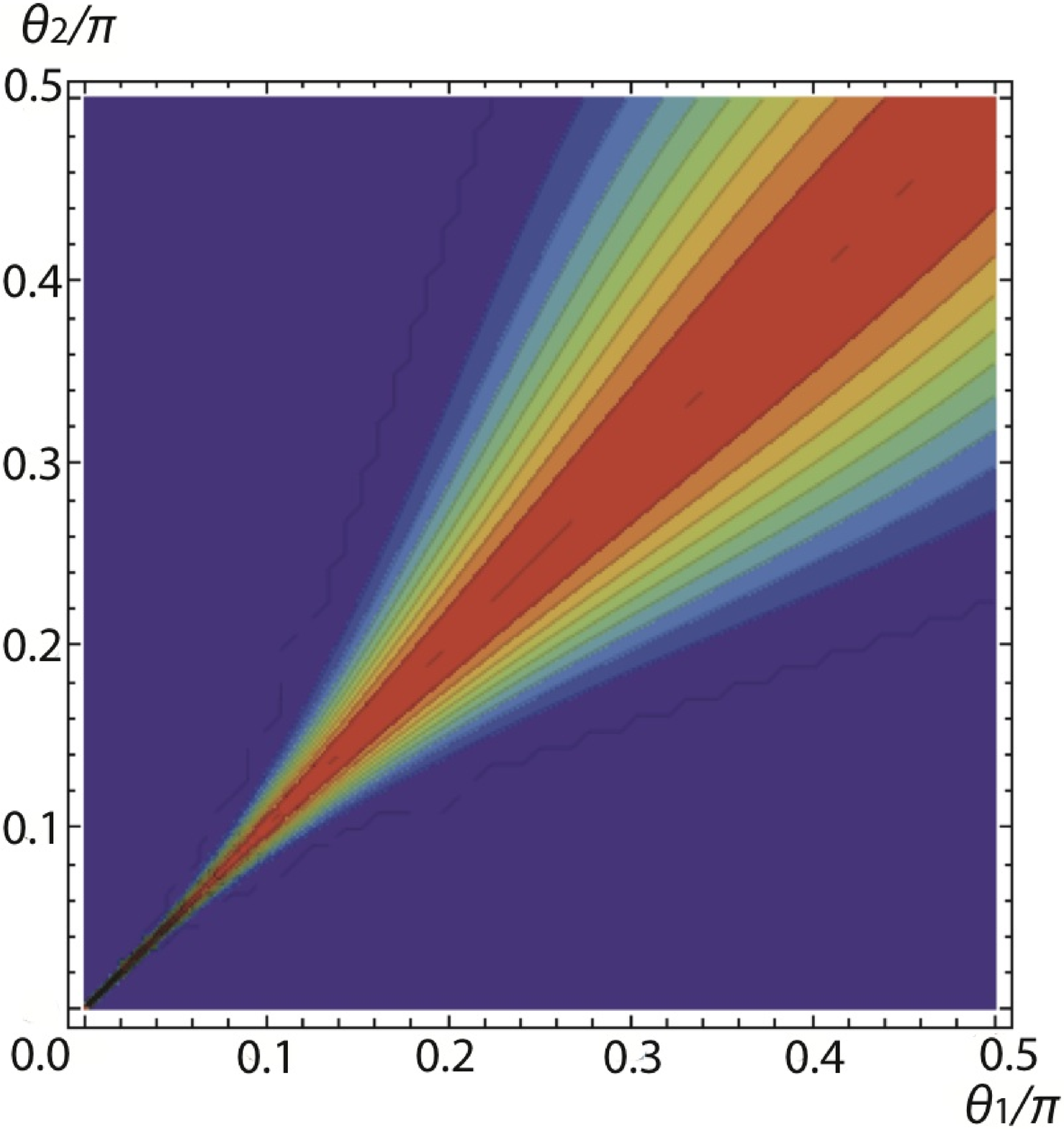}}
\end{array}$
  \includegraphics[width=4in]{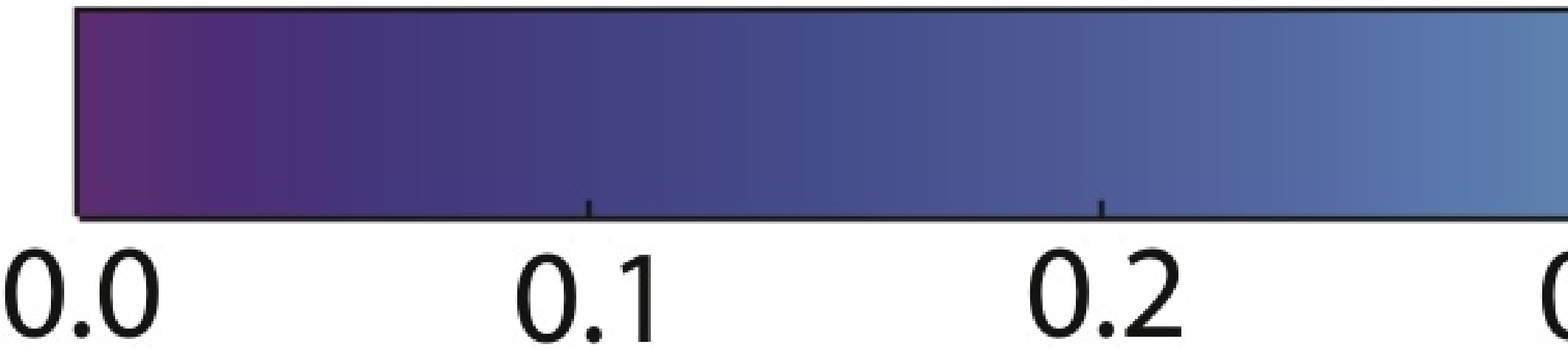}
\caption{Contour plots of (a) the fidelity between the attractor $\rho_{1}$ and the $\ket{T_{0}} \bra{T_{0}}$ state; and (b) the entanglement of formation $E_{F}$ of $\rho_{1}$; each are plotted as a function of $\theta_{1}$ and $\theta_{2}$ $\in (0, \pi/2]$.}
\label{fig:fident}
\end{center}
\end{figure}

Now, we can find how close $\rho_{1}$ is to the $\ket{T_{0}} \bra{T_{0}}$ state, for various values of $\theta_{1}$ and $\theta_{2}$, by calculating the fidelity $F$ as defined in~\cite{jozsa94}. In our case, we have
\begin{equation}
F(\rho_{1}, \ket{T_{0}} \bra{T_{0}}) = \Tr \bigg( \sqrt{\sqrt{\rho_1} \ket{T_0} \bra{T_0} \sqrt{\rho_1}} \ \bigg) = \sqrt{\frac{b + \frac{1}{\sqrt{2}}}{2(a + b)}}
\end{equation}
which takes a value of unity when $\theta_{1} = \theta_{2}$, as expected. Its contour plot in Fig.~\ref{fig:fidenta} illustrates that even when $\theta_{1}$ and $\theta_{2}$ are different, $\rho_{1}$ is still very close to the $\ket{T_{0}} \bra{T_{0}}$ state for any $(\theta_{1}, \theta_{2})$ in the central region. The fidelity values also indicate the levels of degradation in our entangled resource through unequal coupling, the degree of which can be further established by calculating the Entanglement of Formation~\cite{wootters98} $E_{F}$ that the bipartite state $\rho_{1}$ has~\cite{munro01}. We construct the contour plot for $E_{F}$ in Fig.~\ref{fig:fidentb} showing that the degree of entanglement $\rho_{1}$ possesses is very large, higher than 0.9, for any $(\theta_{1}, \theta_{2})$ in the central red region.

\begin{figure}[h]
\begin{center}$
\begin{array}{ccc}
\hspace{-0.8cm}  \subfigure[$ $ Condition]{\label{fig:percentage} \includegraphics[width=2.2in]{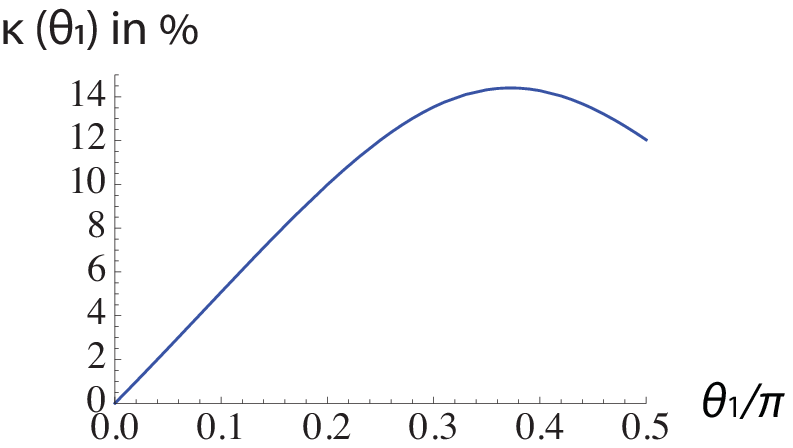}} &
  \subfigure[$ $ random $\theta_{i} \in (1.2, 1.4)$]{\label{fig:1214} \includegraphics[width=1.8in]{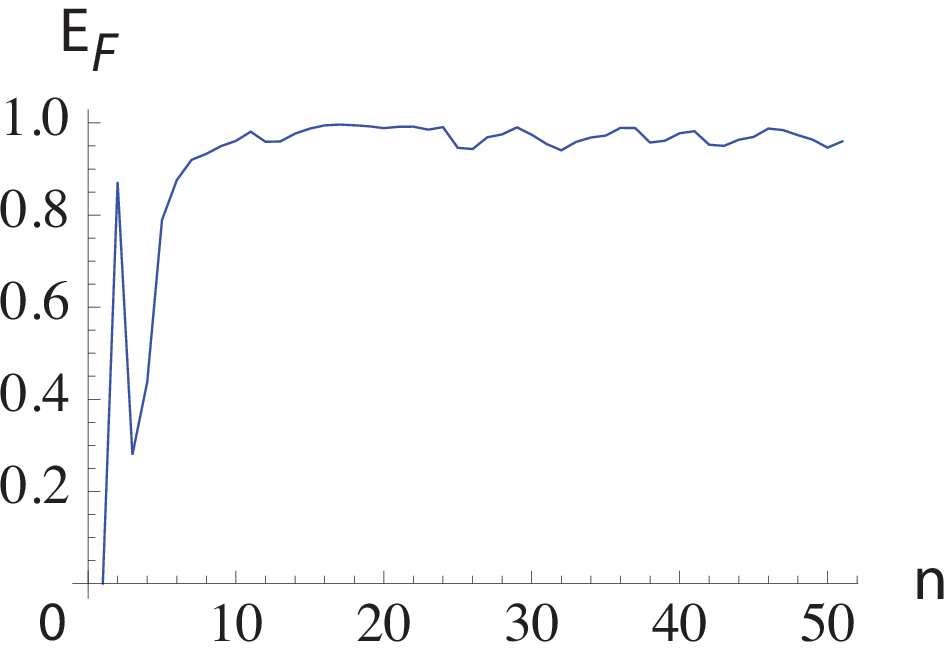}} &
  \subfigure[$ $ random $\theta_{i} \in (0.1, 0.103)$]{\label{fig:010103} \includegraphics[width=2.1in]{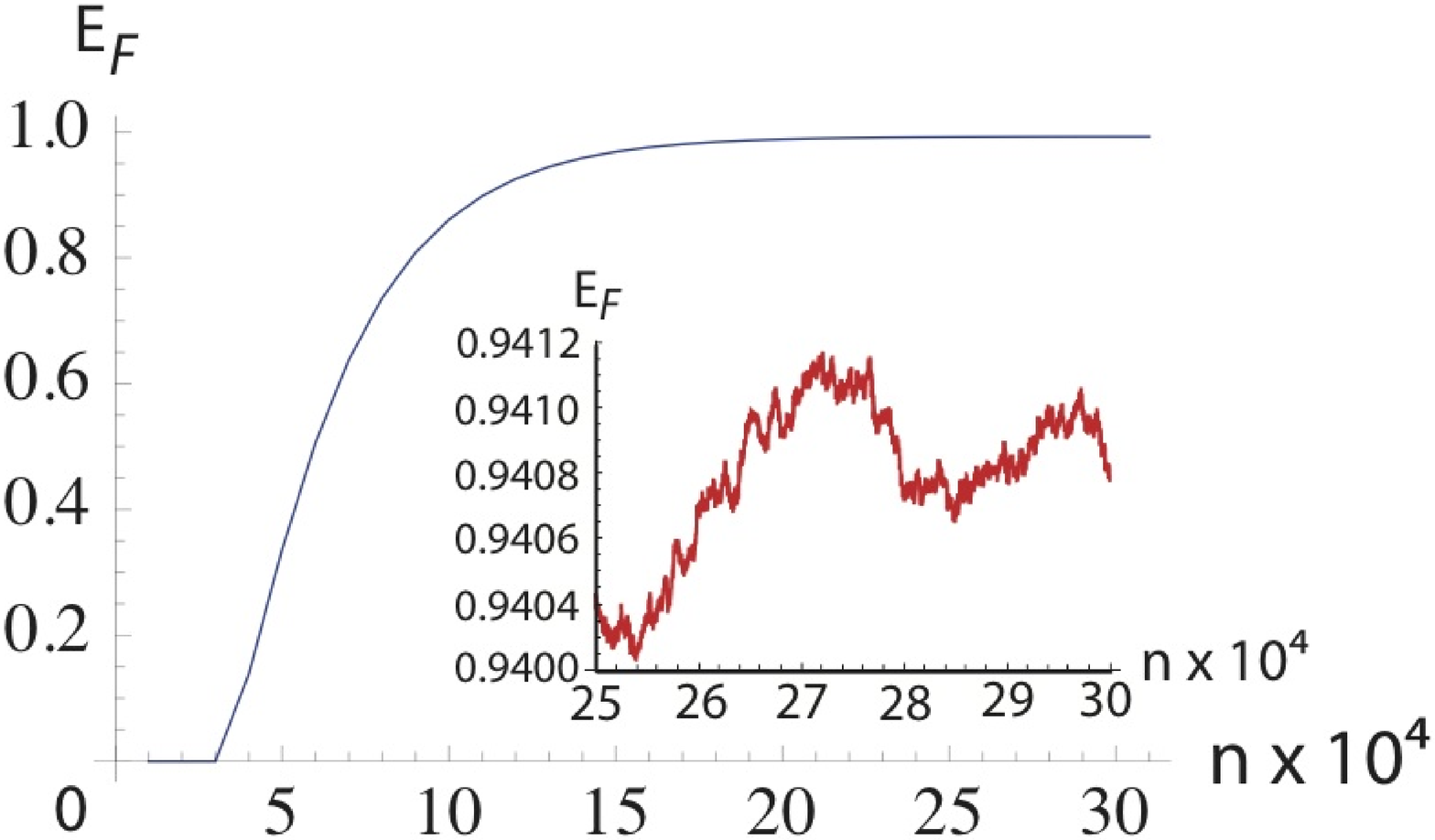}}
\end{array}$
\caption{(a) Plot of the maximum allowed percentage difference $\kappa (\theta_{1})$ against $\theta_{1}/\pi$ for obtaining final $E_{F} (\rho_{s}) \geq 0.9$; (b) \& (c) Example plots of the evolutions of $E_{F}$ for the static spins as a function of the number of flying pairs $n$, for strong and weak random couplings, respectively. The inset in (c) zooms in on the corresponding fluctuations.}
\label{fig:asym}
\end{center}
\end{figure}

The attractor state $\rho_{1}$ remains similar for $(\theta_{1}, \theta_{2})$ values in the same colour region in Fig.~\ref{fig:fident}, and close to $\ket{T_{0}}\bra{T_{0}}$ for the central red region. As a result, a high degree of entanglement with $E_{F} \geq 0.9$ is achieved for the static spins even when $\theta_{i} (k)$ varies with the iteration $k$, as long as in each round $\Delta \theta = |\theta_{1} - \theta_{2}|$ satisfy the condition set in Fig.~\ref{fig:fidentb} such that $(\theta_{1}, \theta_{2})$ lies in the central red region. The maximum possible percentage difference in coupling strengths needed to maintain a highly entangled state may be defined as $\kappa (\theta_{1}) := \max_{E_{F} \geq 0.9} (\Delta \theta/ \theta_{1})$, and this quantity depends on $\theta_{i}$ itself, as shown in Fig.~\ref{fig:percentage}. The maximum value for $\kappa (\theta_{1})$ occurs at a $\theta_{1}$ value away from $\pi/2$ because the maximum allowed $\Delta \theta$ values for final $E_{F} (\rho_{s}) \geq 0.9$ do not vary much for $\theta_{i}$ close to $\pi/2$. For $\theta_{i}$ as weak as 0.1, the maximum allowed percentage difference $\kappa (0.1) \simeq 1.6 \%$. We show two example plots for the evolutions of $E_{F}$ for the static spins vs $n$ for $\theta_{i}$ values that can vary within some range from round to round, in Fig.~\ref{fig:1214}-(c). From these results we see that the convergence times are of the same order as for the case when $\theta_{1} = \theta_{2}$ (see Fig.~2) for both strong and weak couplings. We can also see that the condition on $\kappa (\theta_{1})$ in each round as specified in Fig.~\ref{fig:percentage} only needs to be satisfied by the majority of the interaction rounds.

\section{Decoherence}

In practice, the environment interacts weakly with both static and flying spins, causing their decoherence. This will degrade our final entangled resource to some extent, and we shall consider such effects in this section.

\subsection{Static}

We first take into account the decoherence of the static spin pair, which in general will be coupled to the surrounding environment. The process can be modelled by a Markovian master equation~\cite{breuer02} with decoherence channels $k$ and corresponding decay rates $\gamma_{k}$:
\begin{equation}
\frac{d}{dt} \rho_{s}(t) = \sum_{k=1}^{15} \gamma_{k} (A_{k} \rho_{s} A_{k}^{\dag} - \frac{1}{2} A_{k}^{\dag} A_{k} \rho_{s} - \frac{1}{2} \rho_{s} A_{k}^{\dag} A_{k})
\end{equation}
where we have ignored the unitary term  $-i [H, \rho_{s}]$ on the assumption that the duration of the spin-spin interaction is much shorter than time interval $t_0$ between successive flying qubits. 

This assumption about the much shorter timescale also means that we do not need to consider decoherence during interactions events. Therefore, the evolution of $\rho_{s}(t)$ can be tracked through application of the map $\tilde{L_{s}}$ to describe interaction events and using the solution of the master equation between events. The resulting behaviour is a function of the products $\gamma_{k} t_{0}$, and we assume $t_{0}$ to be constant for simplicity. 

For certain decoherence models, analytical results can be found. For example, if we assume that there are two independent dephasing channels $\sigma_{z}^{s_{1}} \bigotimes I^{s_{2}}$ and $I^{s_{1}} \bigotimes \sigma_{z}^{s_{2}}$ with rates $\gamma_{1}$ and $\gamma_{2}$ respectively, then for $\theta_{1} = \theta_{2}$ the final fidelity of the state of the static spins with respect to $\ket{T_{0}}$ is, immediately prior to an interaction events:
\begin{equation}
\sqrt{\frac{2 + 2 e^{2 (\gamma_{1} + \gamma_{2}) t_{0}} \csc^{4} \theta - \cot^{2} \theta (2 + 2 \csc^{2} \theta)}{4e^{2 (\gamma_{1} + \gamma_{2}) t_{0}} \csc^{2} \theta (2 \csc^{2} \theta - 1) - 4\cot^{2} \theta (1 + 2 \csc^{2} \theta)}}
\end{equation}

\begin{figure}[h]
\begin{center}
\includegraphics[width=4in]{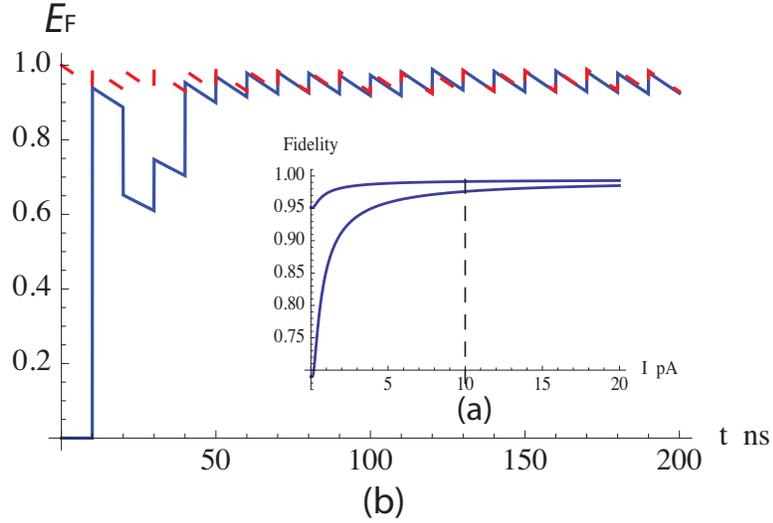}
\caption{Plots of the state of the static spin pair in the presence of pure dephasing errors at random $\theta_{i} \in [1.3, 1.4]$ and $\gamma_{1} = \gamma_{2} = 1$ MHz: (a) Fidelity of final states at equilibrium against current with the higher and lower lines representing the states right after and before XY couplings, respectively; (b) $E_{F}$ against time for a spin-entangled current of $10$ pA, with the dashed and solid lines representing the initial static spins being in the $\ket{T_{0}}$ and the pure $\ket{\uparrow\uparrow}$ states, respectively. For the latter case, the initial rise and the subsequent fall in $E_{F}$ is due to the presence of the singlet state.}
\label{fig:deco}
\end{center}
\end{figure}

We know that for the device to generate a large degree of static entanglement, we need $\theta_{1}$ and $\theta_{2}$ to be in the central red region of Fig.~\ref{fig:fident}. For example, working with random $\theta_{i} \in [1.3, 1.4]$, calculations show that robustness to decoherence requires that $\gamma_{k} t_{0}$ are at most on the order of $10^{-2}$. For $\gamma_{1} = \gamma_{2} = 1$ MHz (i.e., a coherence time of 1 $\mu$s), Fig.~\ref{fig:deco}(a) illustrates the behaviour of the fidelity against current of flying pairs $\frac{e}{t_{0}}$, with $e$ the electronic charge. When the current is large enough, more than $10$ pA (i.e., $t_{0} \lesssim 10$ ns) in this case, the fidelity only fluctuates slightly. Fig.~\ref{fig:deco}(b) then shows that the final state of the static spin pair possesses a large degree of entanglement and does not depend on the initial set up. A similar effect is observed for the bit-flip errors, with slightly reduced final fidelity. This 10 pA is much smaller than the nA scales observed in~\cite{hofstetter09, herrmann10}, i.e. the strong coupling regime is robust to decoherence of the static spins. For weak couplings, a higher current is required for the same decoherence rates, e.g., at $\sim$ 100 nA ($t_{0} \sim 1$ ps) is needed for $\theta_{i} \simeq 0.03$ (or at $\sim$ 1 nA with spin coherence times in~\cite{bluhm10}). We see that decoherence will reduce the fidelity and hence the degree of entanglement of the steady state for the static spins, and thus in Fig.~\ref{fig:fident} the regions for high fidelity and high degree of entanglement will be narrower in the presence of decoherence.

\subsection{Flying} \label{sec:fd}

Next, let us consider the decoherence of the flying singlets before they interact with the static spins. The whole analysis in Sec.~\ref{sec:analysis} applies when we refine our model by allowing random error deviations $\epsilon^{(k)}$ of $\rho_{f}^{(k)}$ from the singlet state in each round such that
\begin{equation}
\rho_{f}^{(k)} = (1 - \epsilon^{(k)}) \ket{S}_{f} \bra{S} + \sum_{j =1}^{3} \epsilon_{j}^{(k)} \sigma_{j}^{f_{1}} \ket{S}_{f} \bra{S} \sigma_{j}^{f_{1}},
\end{equation}
where the errors $\epsilon^{(k)} = \sum_{j = 1}^{3} \epsilon_{j}^{(k)}$ for each $k$ are due to spin-orbit coupling and interactions with the environment as the flying qubit propagates. The $\sigma_{j}$'s are the Pauli $\sigma_{1,2,3}\equiv\sigma_{x, y, z}$ matrices that correspond to different errors on one of the two flying qubits. We find that the single attractor state now has small error-dependent terms $c$ for the corner entries on the minor diagonal of $\rho_{1}$ in Eq.~\ref{eq:evecp}, and both $a$ and $b$ now also depend on $\epsilon_{j}^{(k)}$. As a result, the final $E_{F} (\rho_{1})$ is reduced, and the larger the errors the smaller it will be. We also find that the strong coupling regime can tolerate larger errors compared with weak coupling. This is because the static entanglement built up per round of interaction is much smaller for weak couplings and the errors $\epsilon_{j}^{(k)}$ can drastically reduce the accumulated entanglement. In either case, the average of $\epsilon^{(k)}$ mainly determines the final $E_{F}$ for the static spins for fixed $\theta_{i}$. For strong couplings, the error tolerance on $\epsilon^{(k)}$ is of order 0.01, which corresponds to flying pairs with an average $E_{F}$ of $\sim 0.8 - 0.9$. This results in a travelling distance on the millimetre scale for the flying spins at the Fermi velocity with typical coherence time of microseconds, before their interactions with the static pair. This error tolerance is much smaller for weak couplings, but this could also be feasibile experimentally since couplings become weaker at high carrier speed, which also means a smaller interaction of the flying qubits with the environment before they arrive at the static qubit sites.

Therefore, for a given separation between the static spins, e.g., 1 mm ($\gg$ 1 $\mu$m, of conventional separations~\cite{recher01, recher02, bena02}), the time it takes the flying electrons (travelling at $\sim 10^{5}$ m$s^{-1}$) to arrive at the static qubit sites is 10 ns, much less than a typical spin coherence time ($\mu$s). Separations of centimetre scales can be achieved for longer spin coherence times~\cite{bluhm10}.

\section{Splitting Efficiency}

\begin{figure}[h]
\begin{center}$
\begin{array}{ccc}
\hspace{-0.2cm}  \subfigure[$ $ $\rho_{1_{s}}$ after one round of unsuccessful splittings]{\label{fig:reduction} \includegraphics[width=1.9in]{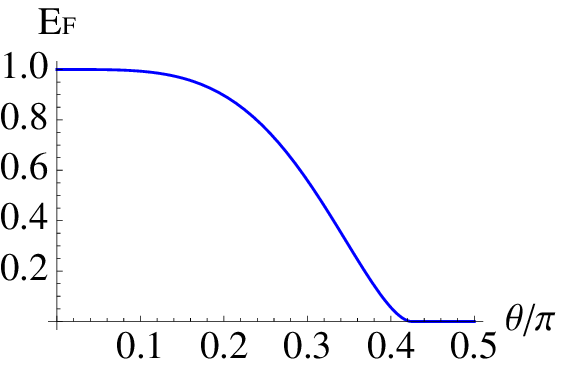}} &
  \subfigure[$ $ $(\theta_{1}, \theta_{2}) = (1.2, 1.3)$ \& $\eta = 95\%$]{\label{fig:strong} \includegraphics[width=1.75in]{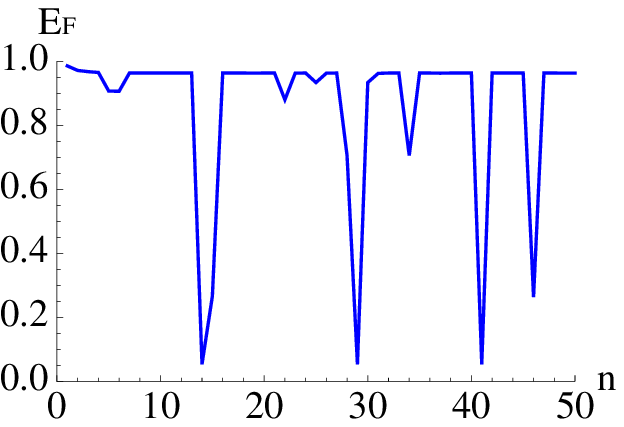}} &
  \subfigure[$ $ $(\theta_{1}, \theta_{2}) = (0.1, 0.1)$ \& $\eta = 99\%$]{\label{fig:weak} \includegraphics[width=2.2in]{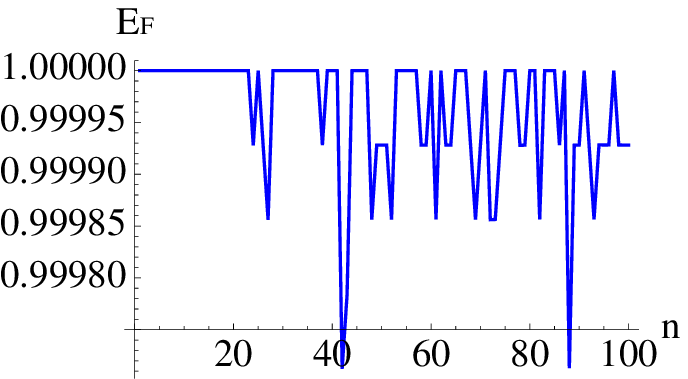}}
\end{array}$
\caption{Plots for the evolutions of $E_{F}$ for the static spins: (a) of the attractor state $\rho_{1}$ after one round of interactions with the flying singlet pair entering the same lead, against various $\theta \in [0, \frac{\pi}{2}]$; (b)-(c) examples of the corresponding steady states against number of rounds $n$ when no other errors are present. Once the steady state is achieved, the success rates required could be lowered.}
\end{center}
\end{figure}

We now consider the efficiency of successful splittings of the Cooper pairs via the double dots in the generator~\cite{hofstetter09,herrmann10} in Fig.~1. When unsuccessful, a singlet pair enters the same lead and this has the effect of reducing the static entanglement (see Fig.~\ref{fig:reduction}). For a success rate $\eta$ of 50\% in~\cite{herrmann10}, the static spins converge to  a completely mixed state regardless of the coupling strengths. We therefore classify the usefulness of the entangled current resource depending on the success rates in terms of final static entanglement we could achieve for the steady state of the static spins. For large coupling strengths, $E_{F} (\rho_{s})$ switches between close to 0 and almost 1 from round to round depending on the success of splitting. Thus in this case, a threshold criterion for $\eta$ would be that the state of the static pair spends a certain fraction of its time in states with $E_{F}$ close to 1, as shown in Fig.~\ref{fig:strong}. In that case, the average $E_F \approx 0.88$ for $\eta=$95\%.  In the weak coupling regime, the steady state $E_{F}$ of the static spins is large and fluctuates slightly for sufficiently high success rates, as shown in Fig.~\ref{fig:weak}. The condition is that the amount of reduction in static entanglement by a single unsuccessful event should be smaller than the accumulated entanglement built up from a certain number $N$ of consecutive successful splittings. In this case then $\eta = \frac{N}{N+1}$, and we find for example, when $\theta \simeq 0.1$, for $\eta \simeq 99.9\%$, then very high entanglement can be maintained (see Fig.~\ref{fig:weak}). 

In either case, for the current resource to be useful in terms of our scheme, the success rates for splittings should improve from those in~\cite{hofstetter09, herrmann10} to at least 90\%. Stronger currents will also make the scheme more robust to the various decoherence sources.

\section{Other Models}
The above analysis was adapted to a Heisenberg exchange model by replacing the effective Hamiltonian in Eq.~1 with 
\begin{equation}
H'_{i} = \frac{g_{i}}{2} (\sx^{s_{i}} \sx^{f_{i}} + \sy^{s_{i}} \sy^{f_{i}} + \sz^{s_{i}} \sz^{f_{i}})
\end{equation}

\begin{figure}[h]
\begin{center}$
\begin{array}{cc}
  \subfigure[Fidelity]{\includegraphics[width=2.1in]{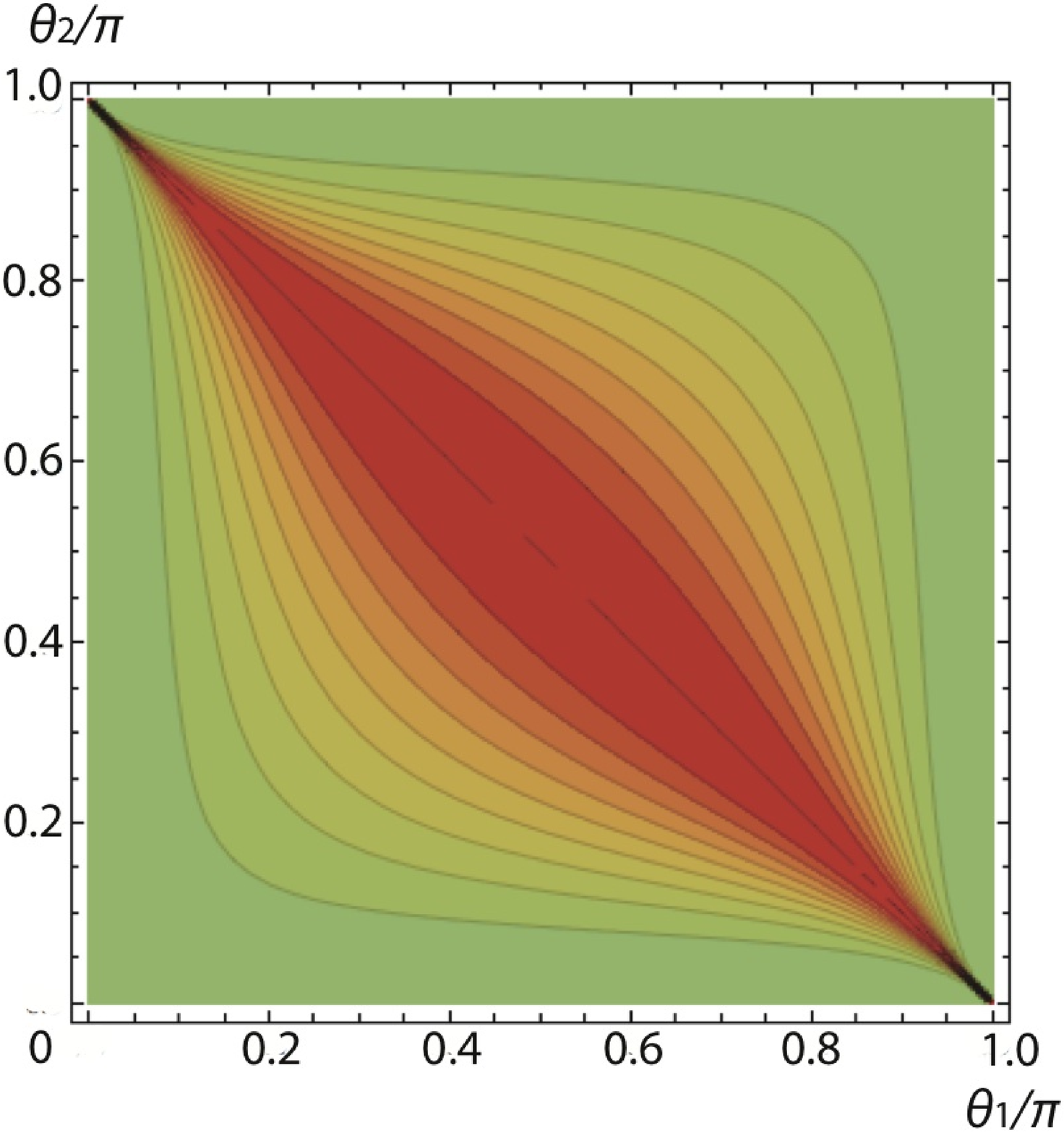}} &
  \subfigure[$E_{F}$]{\includegraphics[width=2.1in]{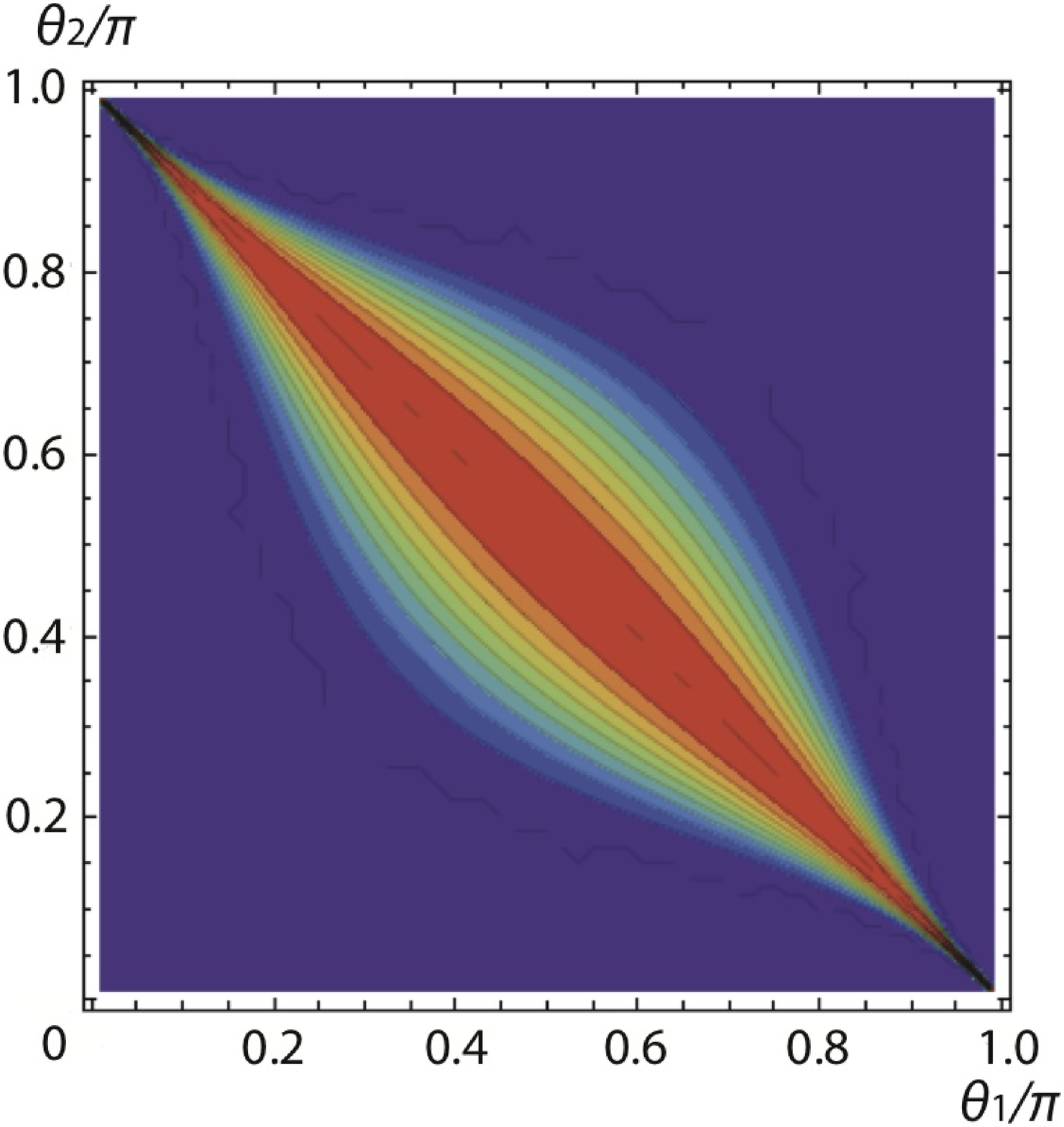}}
\end{array}$
\includegraphics[width=3in]{Figure8.eps}
\caption{Contour plots for the single attractor state of the static spins in the Heisenberg exchange model: (a) Fidelity with respect to the singlet state as a function of $\theta_{i} \in [0, \pi]$; (b) $E_{F}$ as a function of $\theta_{i} \in [0, \pi]$.}
\label{fig:heisenberg}
\end{center}
\end{figure}

We then obtained similar results for the Heisenberg exchange model; the only difference is that this time the singlet state is the only attractor for Eq.~\ref{eq:cpmap}, and the condition of similar coupling strengths is now replaced by the requirement of $\theta_{1} (k) + \theta_{2} (k) \simeq \pi$ in each round of interaction (see Fig.~\ref{fig:heisenberg} in comparison with Fig.~\ref{fig:fident}). Given this understanding, the results on the decoherence effects as well as the splitting efficiency are also similar.

The Heisenberg and XY coupling models can be regarded as examples of a more general anisotropic coupling~\cite{dejongh01}, and both can be realized in suitable materials~\cite{dejongh01, gunlycke06}. For example, the Heisenberg exchange model can effectively describe the spin evolutions of forward scattering electrons: The evolution of the total wavefunction of the scattering electrons is determined by the Schr\"{o}dinger equation with Coulomb repulsion terms, together with the Pauli principle the spin evolution is determined~\cite{gunlycke06}.

\section{Conclusion}

We have shown that distributed and static spin entanglement can be generated from a source of entangled current and weak, passive, Coulomb interactions. The entanglement generated is robust to various error sources.  We therefore anticipate that spin entangled currents can be utilized in the way we have proposed in a wide range of experimental systems.

\section*{Acknowledgments}

B.W.L. acknowledges the Royal Society for a University Research Fellowship. Y. P. thanks Hertford College, Oxford for a scholarship.


\section*{Reference} 
 
\begin{thebibliography}{99}

\bibitem{nielsen00}
M.~A. Nielsen and I.~L. Chuang.
\newblock {\em Quantum Computation and Quantum Information}.
\newblock Cambridge, 2000.

\bibitem{raussendorf01}
R.~Raussendorf and H.~J. Briegel.
\newblock A one-way quantum computer.
\newblock {\em Phys. Rev. Lett.}, 86:5188, 2001.

\bibitem{kok10}
P.~Kok and B.~W. Lovett.
\newblock {\em Introduction to Optical Quantum Information Processing}.
\newblock Cambridge, 2010.  
  
\bibitem{barrett05}
S.~D. Barrett and P.~Kok.
\newblock Efficient high-fidelity quantum computation using matter qubits and linear optics.
\newblock  {\em Phys. Rev. A}, 71:060310, 2006.

\bibitem{bose99}
S.~Bose, P.~L. Knight, M.~B. Plenio, and V.~Vedral.
\newblock Proposal for Teleportation of an Atomic State via Cavity Decay.
\newblock {\em Phys. Rev. Lett.}, 83:5158, 1999.

\bibitem{cabrillo99}
C.~Cabrillo, J.~I. Cirac, P.~Garc\'{i}a-Fern\'{a}ndez, and P.~Zoller.
\newblock Creation of entangled states of distant atoms by interference.
\newblock {\em Phys. Rev. A}, 59:1025, 1999.

\bibitem{lo99}
H.~-K. Lo and H.~F. Chau.
\newblock Unconditional Security Of Quantum Key Distribution Over Arbitrarily Long Distances.
\newblock {\em Science}, 283:2050, 1999.

\bibitem{shor00}
P.~W. Shor and J.~Preskill.
\newblock Simple Proof of Security of the BB84 Quantum Key Distribution Protocol.
\newblock {\em Phys. Rev. Lett.}, 85:441, 2000.
  
\bibitem{bennett93}
C.~H. Bennett, G.~Brassard, C.~Crepeau, R.~Josza, A.~Peres, and W.~K. Wootters.
\newblock Teleporting an unknown quantum state via dual classical and Einstein-Podolsky-Rosen channels.
\newblock {\em Phys. Rev. Lett.}, 70:1895, 1993.
 
\bibitem{saraga03} 
D.~S. Saraga and D.~Loss.
\newblock Spin-entangled currents created by a triple quantum dot.
\newblock {\em Phys. Rev. Lett.}, 90:166803, 2003.
  
\bibitem{oliver02}  
W.~D. Oliver, F.~Yamaguchi, and Y.~Yamamoto.
\newblock Electron Entanglement via a Quantum Dot.
\newblock {\em Phys. Rev. Lett.}, 88:037901, 2002.

\bibitem{kolli09}
A.~Kolli, S.~C. Benjamin, J.~G. Coello, S.~Bose, and B.~W. Lovett.
\newblock Large spin entangled current from a passive device.
\newblock {\em New. J. Phys.}, 11:013018, 2009.

\bibitem{hofstetter09}
L.~Hofstetter, S.~Csonka, J.~Nygard, and C.~Schoenenberger.
\newblock Cooper pair splitter realized in a two-quantum-dot Y-junction.
\newblock {\em Nature}, 461:960, 2009.

\bibitem{herrmann10}
L.~G. Herrmann, F.~Portier, P.~Roche, A.~L. Yeyati, T.~Kontos, and C.~Strunk.
\newblock Carbon Nanotubes as Cooper-Pair Beam Splitters.
\newblock {\em Phys. Rev. Lett.}, 104:026801, 2010.

\bibitem{recher01}
P.~Recher, E.~V. Sukhorukov, and D.~Loss.
\newblock Andreev tunneling, Coulomb blockade, and resonant transport of nonlocal spin-entangled electrons.
\newblock {\em Phys. Rev. B}, 63:165314, 2001.

\bibitem{recher02} 
P.~Recher and D.~Loss.
\newblock Superconductor coupled to two Luttinger liquids as an entangler for electron spins.
\newblock {\em Phys. Rev. B}, 65:165327, 2002.

\bibitem{bena02}
C.~Bena, S.~Vishveshwara, L.~Balents, and M.~P. A. Fisher.
\newblock Quantum Entanglement in Carbon Nanotubes.
\newblock {\em Phys. Rev. Lett.}, 89:037901, 2002.

\bibitem{watt08}
A.~A. R. Watt, M.~R. Sambrook, S.~V. Burlakov, K.~Porfyrakis, and G.~A. D. Briggs.
\newblock Probing the interior environment of carbon nano-test-tubes.  
\newblock {\em arXiv:0810.3124v1}, 2008.

\bibitem{nasibulin07}
A.~G. Nasibulin {\it et al.} 
\newblock A novel hybrid carbon material.
\newblock {\em Nature Nanotechnology}, 2:156, 2007.
  
\bibitem{barnes00}  
C.~H. W. Barnes, J.~M. Shilton, and A.~M. Robinson.
\newblock Quantum computation using electrons trapped by surface acoustic waves.
\newblock {\em Phys. Rev. B}, 62(12):8410, 2000.

\bibitem{bluhm10}
H.~Bluhm, S.~Foletti, I.~Neder, M.~Runder, D.~Mahalu, V.~Umansky, and A.~Yacoby.
\newblock Long coherence of electron spins coupled to a nuclear spin bath.
\newblock {\em arXiv:1005.2995v1}, 2010.
  
\bibitem{ge08}
L.~Ge, B.~Montanari, J.~H. Jefferson, D.~G. Pettifor, N.~M. Harrison, and G.~A. D. Briggs.
\newblock Modeling spin interactions in carbon peapods using a hybrid density functional theory.
\newblock {\em Phys. Rev. B}, 77:235416, 2008.

\bibitem{jozsa94}
R.~Jozsa.
\newblock Fidelity for mixed quantum states.
\newblock {\em J. Mod. Optic.}, 41:2315, 1994.

\bibitem{wootters98}
W.~K. Wootters,
\newblock Entanglement of Formation of an Arbitrary State of Two Qubits.
\newblock {\em Phys. Rev. Lett.}, 80:2245, 1998.
  
\bibitem{munro01}
W.~J. Munro, D.~F. V. James, A.~G. White, and P.~G. Kwiat.
\newblock Maximizing the Entanglement of Two Mixed Qubits.
\newblock {\em Phys. Rev. A}, 64:030302, 2001.
   
\bibitem{breuer02}
H.~-P. Breuer and F.~Petruccione.
\newblock {\em The Theory of Open Quantum Systems}.
\newblock Oxford, 2002.  
  
\bibitem{dejongh01}
L.~J. De Jongh and A.~R. Miedema.
\newblock Experiments on simple magnetic model systems.
\newblock {\em Adv. Phys.}, 50:947, 2001.
 
\bibitem{gunlycke06}
D.~Gunlycke, J.~H. Jefferson, T.~Rejec, A.~Ram\u{s}ak, D.~G. Pettifor, and G.~A. D. Briggs.
\newblock Entanglement between static and flying qubits in a semiconducting carbon nanotube.
\newblock {\em J. Phys.: Condens. Matter}, 18:S851, 2006.

\end{thebibliography}
\end{document}